# THE MODEL OF NEUTRINO VACUUM FLAVOUR OSCILLATIONS AND QUANTUM MECHANICS

Boris I. Goryachev

D.V. Skobeltsyn Institute of Nuclear Physics, M.V. Lomonosov Moscow State University, Moscow, Russian Federation

E-mail: bigor@srd.sinp.msu.ru

It is shown that the model of vacuum flavour oscillations is in disagreement with quantum mechanics theorems and postulates. Features of the model are analyzed. It is noted that apart from the number of mixed mass states neutrino oscillations are forbidden by Fock-Krylov theorem. A possible reason of oscillation model inadequacy is discussed.

## 1. Introduction

For a number of decades search for neutrino oscillations is the basic direction in the field of massive neutrino experimental physics.

Idea of these oscillations caused by transitions $\nu \leftrightarrows \bar{\nu}$ was advanced on the analogy with $K^0 \overline{K^0}$ oscillations [1] and later was extended to oscillations of neutrino with different flavours [2, 3].

This type of oscillations is of vital importance for the attempts to explain solar neutrino deficit, and almost all experiments in this field were directed to the search for flavour neutrino oscillations. Hereinafter precisely these oscillations will be regarded as neutrino oscillations.

Neutrino oscillations *hypothesis* is based on the assumption that flavour neutrino states $|\nu_f\rangle$ being weak interactions Hamiltonian eigenstates, are not eigenstates $|\nu_i\rangle$ of mass operator and can be obtained by mixing the latter:

$$|\nu_f\rangle = \sum_i U_{fi}|\nu_i\rangle, \qquad (1)$$

where $U$ – unitary mixing matrix.

The number of mixed mass states in (1) is equal to the number of interaction states. Usually two variants: 2*f*- and 3*f*-oscillations, are used for the experimental data analysis. The latter variant takes into account the full number of possible flavours (three). As is known mixing (1) results in the transitions from one flavour to another, so observation of neutrino oscillations would mean flavour lepton numbers nonconservation.

In spite of prevailing hopefulness in the evaluation of the possibilities of the neutrino oscillation model, a number of questions related to such evaluation still remains. Foremost two of them should be marked out:

1. Why none of "direct" experiments, which do not use search for neutrino oscillations, discovers nonconservation of the lepton flavour number?
2. Why does mixing in quark sector is significantly smaller than supposed mixing of neutrino states?

In the majority of the studies parameters of the neutrino oscillations models are evaluated according to the existed experimental data. This way is associated with considerable difficulties. In particular, the events, qualified by experimenters as desired ones, can be simulated by background processes. Therefore the problem of studying of the neutrino oscillations model basing on the quantum mechanics theorems, not taking into account experimental data, is of particular interest.

The present report concerns precisely this problem.
The analysis below is based on the theorems true for isolated systems. Therefore, conclusions obtained in the present study concern vacuum neutrino oscillations.

Effects of the interaction with matter [4, 5] are out of the question.
As usually, neutrino are considered as stable particles.

Section 2 includes some specific characteristics of the vacuum neutrino oscillations model (in 2*f*- and 3*f*-variants), which lead to the limits for the selection of the model parameters number.

Consistency of the oscillation model and Fock-Krylov theorem [6] along with specific characters of the missing of neutrino mass states are analyzed in Section 3. Appendix includes the notations for the standard mixing matrix elements, used in the present report due to their spelling convenience.

## 2. Specific characteristics of the vacuum neutrino oscillations model

### 2.1. *2-f oscillations*

In the case of two flavours we'll designated interaction state, generated in the source at $t = 0$ as $|\nu_a\rangle$, and the state missing from the initial neutrino flux at $t = 0$ as $|\nu_b\rangle$.

In respect to the solar neutrino $|\nu_a\rangle \equiv \nu_e$, and muon neutrino $\nu_\mu$ can occur as $|\nu_b\rangle$.

For the long-baseline accelerator experiments the following identification is possible: $|\nu_a\rangle \equiv \nu_\mu$ and $|\nu_b\rangle \equiv \nu_\tau$.

Then according to (1) and agreed notations (see Appendix)

$$|\nu_a\rangle = \cos\theta |\nu_1\rangle + \sin\theta |\nu_2\rangle \qquad (2a)$$

$$|\nu_b\rangle = -\sin\theta |\nu_1\rangle + \cos\theta |\nu_2\rangle, \qquad (2b)$$

where θ is mixing angle. Mass states $|v_i\rangle$ can be expressed in the following manner (hereinafter $\hbar=c=1$)

$$|v_i\rangle = C|v(p_i)\rangle \exp(-i\varepsilon_i t) \qquad (3)$$

where $|v(p_i)\rangle$ – eigenstate of the momentum operator with eigenvalue $p_i$, and $\varepsilon_i$ – total energy of $|v_i\rangle$ state.

Mass states $|v_i\rangle$ are the solutions of the wave equation and are described by plane waves. If neutrino move, for instance, along z-axis, that:

$$|v_i\rangle \sim exp\left[i(p_i z - \sqrt{p_i^2 + m_i^2}\,t)\right] \qquad (4)$$

As applied to the experiments on the search for neutrino oscillations $t$ value means time of neutrino transit from the source to the detector.

As usually, wave functions $|v(p_i)\rangle$ are normalized in the appropriate volume and system of these functions is orthonormal. So if $p_i$ are different, during calculation of matrix elements cross terms formed by not coincident mass states will become zero, and it will lead to zero oscillations. As is known these oscillations appear due to interference of different mass states $|v_i\rangle$ in (1) (for instance, see [7]) and restriction

$$p_i = p \qquad (5)$$

is the necessary criterion for the neutrino vacuum oscillations model. Selection of $C$-constant in (3) allows to provide "correct" initial conditions concerning intensities of the fluxes of neutrino with different flavours.

We'll express the norms $N_f$ of $|v_a\rangle$ and $|v_b\rangle$ states and means $\langle F \rangle$ of physical quantities $F$ for these states according to the rules

$$N_f = \langle v_f | v_f \rangle \quad (f=a,b) \qquad (6)$$

and

$$\langle F \rangle = \langle v_f | \hat{F} | v_f \rangle / \langle v_f | v_f \rangle \quad (f=a,b), \qquad (7)$$

where $\hat{F}$ is an operator of the given quantity $F$.

For $N_f(t)$ the following formulas are true:

$$N_f(t) = C^2[1 \pm \sin(2\theta)\cos(\omega t)], \qquad (8)$$

where plus and minus signs are attributed to $a$- and $b$-states, respectively. In the case of relativistic neutrino $(p \gg m_i)$

$$\omega \equiv \omega_r = (m_1^2 - m_2^2)/2p \qquad (9)$$

According to (5), (7) and (8) mean value of momentum ($\hat{F} = -i\partial/\partial z$) is given by the expression

$$\langle p \rangle = p, \qquad (10)$$

came out from (5). For the problem at hand momentum is a motion integral.

For the calculation of the mean values of energy $\langle E \rangle$ (for free particles $\hat{F} = i\partial/\partial t$) condition (5), which leads to the appearance of cross terms, is responsible for complex value of energy

$$\langle E \rangle = Re\langle E \rangle + iIm\langle E \rangle \qquad (11)$$

for any time $t$ (except $t = 0, \frac{\pi}{\omega}, 2\pi/\omega$, etc.; these values are excluded from consideration up to Section 3).

For relativistic neutrino

$$E_0 \equiv Re\langle E \rangle = p + \frac{1}{2p} \left\{ \frac{m_1^2 + m_2^2}{2} \pm \frac{m_1^2 - m_2^2}{2} \cos(2\theta) [1 \pm \sin(2\theta) \cos(\omega_r t)]^{-1} \right\} \qquad (12)$$

and

$$Im\langle E \rangle = \mp \frac{\sin(2\theta)}{4p} (m_1^2 - m_2^2) \sin(\omega_r t)[1 \pm \sin(2\theta) \cos(\omega_r t)]^{-1} \qquad (13)$$

For plus-minus signs the upper and lower signs are attributed to $a$- and $b$-states, respectively. The same note is true for the formulas (15), (8a), (13a), (19)-(21) along with (19a) and (21a).

Complex energy $\langle E \rangle$ embodies the continuity the energy $E$ distribution for the flavour states at hand (1).

We'll show that $Im\langle E \rangle$ value is equal to the half-width $\Gamma_f$ (in terms of spectroscopy) of $|\nu_f\rangle$ states, which is caused by transitions between $a$- and $b$-states. Let's analyze the number of flavour states, proportional to $N_f$ in (8). Then according to definition from [8]

$$N_f \Gamma_f = \{number\ of\ transitions\ per\ unit\ time\} \equiv dN_f/dt \qquad (14)$$

Indeed, taking into consideration (8), (9) and (14) we obtain

$$\Gamma_f = (dN_f/dt)/N_f = \mp \frac{\sin(2\theta)}{2p} (m_1^2 - m_2^2) \sin(\omega_r t)[1 \pm \sin(2\theta) \cos(\omega_r t)]^{-1} \qquad (15)$$

and comparing (15) and (13), we have

$$Im\langle E \rangle = \frac{1}{2} \Gamma_f \qquad (16)$$

$\Gamma_f$ width of state points to the measure of uncertainty for the value of energy (mass) of the flavour neutrino. Energy $E_0 \equiv Re\langle E \rangle$ defines "center of gravity" of the energy distribution of neutrino and can be identified with the mean energy of this distribution. Hereinafter we'll consider mean energy of the flavour neutrino as $E_0$ value, taking into account that for the problem at hand these neutrino are emitted with the same momentum $p$.

For the oscillation model $\Gamma_f$ width also oscillates, because as can be seen from (8) and (14) during different time periods $t$ each state $|\nu_a\rangle$ and $|\nu_b\rangle$ is mostly either "filled" or "cleaned".

Usually during parametrization of the experimental data within the frames of the oscillation model and *2f*-variant two parameters are found – mixing angle $\theta$ and value $(m_1^2 - m_2^2)$.

But this number of parameters is excess.

Really, formula (12) shows, that $E_0$ varies with time, while quantum mechanics canons require mean energy to be conserved in the isolated system. Fixing of $E_0$ is possible by setting $\theta = \pi/4$, which means that condition of maximal mixing of neutrino mass states is realized.

In the case of maximal mixing formulas (8), (12) and (13) are simplified in the following manner:

$$N_f(t) = C^2[1 \pm \cos(\omega_r t)], \tag{8a}$$

$$E_0 = p + \frac{1}{2p}\frac{m_1^2 + m_2^2}{2} \tag{12a}$$

and

$$Im\langle E \rangle = \mp \frac{m_1^2 - m_2^2}{4p} \sin(\omega_r t)[1 \pm \cos(\omega_r t)]^{-1}. \tag{13a}$$

Besides maximal mixing in order to obtain the "correct" initial conditions

$$N_a(0) = 1 \text{ and } N_b(0) = 0 \tag{17}$$

according to (8a) it is necessary to set

$$C = 1/\sqrt{2}. \tag{18}$$

Conditions (17) mean, that at *t=0* only *a*-states should be "filled". Due to relation (5) the present problem can be analyzed in the intrinsic frame of reference of neutrino (*p=0*). Thereby relations (8), (12) and (13) lead to the following:

$$N_f(t) = C^2[1 \pm \sin(2\theta)\cos(\omega_0 t)], \ C = 1/\sqrt{2}, \tag{19}$$

$$E_0 = \frac{(m_1 + m_2)}{2} + \frac{m_1 - m_2}{2}\cos(2\theta)[1 \pm \sin(2\theta)\cos(\omega_0 t)]^{-1}, \tag{20}$$

$$Im\langle E\rangle = \mp \frac{m_1-m_2}{2}\sin(2\theta)\sin(\omega_0 t)[1 \pm \sin(2\theta)\cos(\omega_0 t)]^{-1}, \quad (21)$$

where

$$\omega_0 = m_1 - m_2 \quad (22)$$

It should be stressed that in any frame of axes the vacuum oscillations model should enforce two conditions:

*(i)* conservation of the mean energy of the free particles (neutrino), which leads to the necessity for the maximal mixing in (1);

*(ii)* initial conditions, which mean that only one type of neutrino should occur in frames of flavour states at *t=0*.

Fulfillment of the condition *(ii)* is taken out in *2f*-variant by means of selection of normalization constant (18) and by meeting *(i)* requirement.

Taking into account *(i)* and *(ii)* in the intrinsic frames of reference of neutrino we obtain

$$N_f(t) = C^2[1 \pm \cos(\omega_0 t)], \quad C = 1/\sqrt{2}, \quad (19a)$$

$$E_0 = \frac{m_1+m_2}{2}, \quad (20a)$$

$$Im\langle E\rangle = \mp \frac{m_1-m_2}{2}\sin(\omega_0 t)[1 \pm \cos(\omega_0 t)]^{-1}. \quad (21a)$$

As seen from (8a) and (19a) meeting the requirements *(i)* and *(ii)* leads to the time dependence of $N_f(t)$, coincident with harmonic oscillations. So the model of the vacuum flavor oscillations is associated with availability of some neutrino "inner clock", which define the rate of transitions from one flavour state into another. Corresponding cyclic frequencies ω in different moving frames of reference should be related by Lorentz transformations, and it lead to the relation:

$$\omega_0 = \gamma \omega_r, \quad (23)$$

where *γ* is Lorentz factor of relativistic neutrino. According to (9) and (22) this relation is realized, if the value *(m₁+m₂)/2*, correspondent to the maximal mixing mode, is taken as effective mass of active neutrino.

During interpretation of experimental data within the frames of oscillation model usually probabilities of the transitions $P(a \rightarrow b; t)$ and $P(a \rightarrow a; t)$, where *t* is time of the neutrino flight from the source to the detector, are calculated. But in the experiment it is possible to measure only the detection probability for neutrino with given flavour. Under this approach it is necessary to know flavour composition of neutrino, generated by the source at *t=0*. Usually this composition is given "manually" on the assumption with lepton numbers' conservation.

In the present report the norms of state vectors $N_f(t)$, which contain all necessary information for comparison with the data of the experiments mentioned above, are calculated. It is easy to show that under realization of *(i)* and *(ii)* requirements the results, obtained by both approaches, are the same.

## 2.2 3f-oscillations

In the most of studies *2f*-variant is used for parametrization of the experimental data within the frames of the oscillation model.

Section 2.2 contains analysis of the problem, to what extent maximal mixing of mass states (i.e. $u_{a1} = u_{a2} = u_{a3}$) allows to enforce natural physical requirements *(i)* and *(ii)* in *3f*-variant. We'll calculate in the neutrino intrinsic frame of reference. Matrix (A.2) will be used as mixing matrix (see Appendix). The following values can be selected:

$$S_2 = 1/\sqrt{3},\ C_2 = \sqrt{2}/\sqrt{3},\ S_3 = C_3 = 1/\sqrt{2}, \tag{24}$$

which lead to the "correct" initial conditions

$$N_a(0) = 1,\ N_b(0) = 0,\ N_c(0) = 0. \tag{25}$$

These conditions do not depend on the selection of $S_1$ (and $C_1$, respectively), if values (24) are fixed. So the form of the mixing matrix in *3f*-variant can vary essentially.

We'll give two examples of these matrices, associated to the maximal mixing of mass states.

1. $S_1 = 0,\qquad C_1 = 1$

$$\begin{aligned}
u_{a1} &= 1/\sqrt{3} & u_{a2} &= 1/\sqrt{3} & u_{a3} &= 1/\sqrt{3} & (26.1)\\
u_{b1} &= -1/\sqrt{2} & u_{b2} &= 1/\sqrt{2} & u_{b3} &= 0 & (26.2)\\
u_{c1} &= -1/\sqrt{6} & u_{c2} &= -1/\sqrt{6} & u_{c3} &= \sqrt{2}/\sqrt{3} & (26.3)
\end{aligned}$$

2. $S_1 = 1/\sqrt{2},\qquad C_1 = 1/\sqrt{2}$

$$\begin{aligned}
u_{a1} &= 1/\sqrt{3} & u_{a2} &= 1/\sqrt{3} & u_{a3} &= 1/\sqrt{3} & (27.1)\\
u_{b1} &= -(\sqrt{3}+1)/2\sqrt{3} & u_{b2} &= (\sqrt{3}-1)/2\sqrt{3} & u_{b3} &= 1/\sqrt{3} & (27.2)\\
u_{c1} &= (\sqrt{3}-1)/2\sqrt{3} & u_{c2} &= -(\sqrt{3}+1)/2\sqrt{3} & u_{c3} &= 1/\sqrt{3} & (27.3)
\end{aligned}$$

Matrices (26) and (27) result in the same expression for the norm $N_a(t)$,

$$N_a(t) = C^2\left\{1 + \tfrac{2}{3}[\cos(m_1 - m_2)t + \cos(m_1 - m_3)t + \cos(m_2 - m_3)t]\right\} \tag{28}$$

$$C = 1/\sqrt{3}, \tag{29}$$

where, as is known, only two values of three $(m_i - m_k)$ are independent. It was already stressed in (25), that $N_a(0) = 1$. Really, it runs from (28) and (29). Here expressions for $N_b(0)$ and $N_c(0)$ are discarded.

Further, for the maximal mixing we obtain the following relation (for $|v_a\rangle$ state)

$$E_0 = \frac{m_1 + m_2 + m_3}{3} +$$

$$\frac{1}{3}\left\{\frac{m_1+m_2+m_3}{3}[\cos(m_1 - m_2)t + \cos(m_1 - m_3)t + \cos(m_2 - m_3)t] - [m_3\cos(m_1 - m_2)t + m_2\cos(m_1 - m_3)t + m_1\cos(m_2 - m_3)t]\right\} \times \left\{1 + \frac{2}{3}[\cos(m_1 - m_2)t + \cos(m_1 - m_3)t + \cos(m_2 - m_3)t]\right\}^{-1} \quad (30)$$

At $t = 0$

$$E_0 = \frac{1}{3}(m_1 + m_2 + m_3) \quad (31)$$

and it corresponds to the value (20a) in the case of *2f*-variant.

The structure of the formula (30) for $E_0$ shows, that this value is not a motion integral and varies with increasing of $t$ (time of the particles' flight from the source to the detector).

So, *3f*-variant in the maximal mixing mode does not allow simultaneous realization of *(i)* and *(ii)* requirements.

Expression for the imaginary component of the $|v_a\rangle$ state's energy in *3f*-variant is as follows:

$$Im\langle E\rangle = -\frac{1}{3}[(m_1 - m_2)\sin(m_1 - m_2)t + (m_1 - m_3)\sin(m_1 - m_3)t + (m_2 - m_3)\sin(m_2 - m_3)t] \times \left\{1 + \frac{2}{3}[\cos(m_1 - m_2)t + \cos(m_1 - m_3)t + \cos(m_2 - m_3)t]\right\}^{-1} \quad (32)$$

As is seen from (32) at $t=0$ the value $Im\langle E\rangle = 0$.

Most often for parametrization of the experimental data mixing matrix of *3f*-variant, reduced to the corresponding matrix of *2f*-variant, is used (see Appendix). As it was shown in Section 2.1, in this case the maximal mixing of mass states allows the realization of both requirements *(i)* and *(ii)*.

### 3. The neutrino oscillations model and the quantum theory canons

3.1   Performed in the previous section analysis is of methodological character. The problem of to what extent the model of the vacuum neutrino oscillations is reasonable within the as regards to the quantum mechanics, is the subject of the present section.

Here it is convenient to apply the theorem by Fock-Krylov [6] (see also [9]). According to this theorem for arbitrary (including quasi-stationary) state $|a\rangle$ of the isolated system the following relation is true:

$$L_a(t) = |\int exp(-iEt)W_a(E)dE|^2 \qquad (33)$$

where $W_a(E)$ is distribution function for energy $E$ in state $|a\rangle$ at $t=0$, and $L_a(t)$ is probability that by the time $t$ the system is still at the $|a\rangle$ state. Function $W_a(E)$ is obviously normalized

$$\int W_a(E)dE = 1 \qquad (34)$$

For the problem at hand the states $|a\rangle$ and $|\nu_a\rangle$ are the same.

Vanishing of the spectroscopic width of the state $|a\rangle$ as time tends to $t=0$ (see relations (13a), (21a), (32)) is the distinctive feature of the oscillation model. As a consequence the following relation is true:

$$W_a(E; t = 0) = \delta(E - E_0), \qquad (35)$$

here $\delta(E - E_0)$ is Dirac delta-function. From (35) it follows that

$$L_a(t) = |exp(-iE_0 t)|^2 = 1 \qquad (36)$$

Formula (36) is true throughout $t$. In other words, initial state $|a\rangle$ does not vary with time. It means that transitions into other states with flavours and, respectively, flavour oscillations are not available.

3.2   The following property can be taken as the crucial point of the oscillation model: mixing states are the states of Hermitean Hamiltonians, but the resulting state, in general, corresponds to non-Hermitean Hamiltonian. Essential points of the problem of neutrino oscillations are convenient to be considered within the intrinsic frame of reference of neutrino. Therefore model's difficulty mentioned above can be qualified as paradox of "mixing by mass", which can be hardly referred to any physical meaning.

The importance of using of Hermitean Hamiltonians in quantum mechanics is well known. In particular, the authors of [10] note that in the field of radioactive decay and scattering of the atomic nuclei and particles all theorems on an expansion of arbitrary function in eigenfunctions, which form a complete system, belong to the set of wave functions for the real eigenvalues of energy $E$. Functions of state, corresponding to complex values of $E$, are excluded from this system of eigenfunctions.

We can ignore the imaginary component of the energy $E$ for long-lived particles and nuclei, but within the frames of the problem of neutrino oscillations this component is essential, because it defines the rate of flavour oscillations.

3.3 It can believed that the mentioned paradox is caused by using of the states superposition principle (see (1)). It is known, that superposition of the states with different values of total electric charge, for instance, are physically unrealized. In the field of quantum theory it is embodied in the rules of superselection [11].

In the field of classical physics the electric charge and the mass of the particle act as charges for electromagnetic and gravitational interactions, respectively.

Besides, for the stable elementary particles, to which neutrino belongs, group-theoretical approach (within the frames of quantum-mechanical Poincare group concept) predicts particular value of the elementary particle mass.

In view of the above notes the model, assuming coincidence of the interaction states and neutrino mass states, has a major appeal in comparison to the oscillation model due to the clearance of the former from the paradox of "mixing by mass". Within the frames of this model lepton numbers of neutrino conserve.

### 4.Conclusions

The model of vacuum flavour oscillations is not in agreement with the theorems and postulates of quantum mechanics.

Even within the frames of two flavours "correct" initial conditions and requirements of neutrino energy to be motion integral is true only in case of maximal mixing of mass states.

Without regard to the number of such states (i.e. to $i$) neutrino oscillations are forbidden by Fock-Krylov theorem.

Although mixing states are described by ordinary plane waves, Hamiltonian of the resulting state is non-Hermitean. This paradox can't be explained within the frames of quantum mechanics canons. But it can be also resolved as neutrino oscillations, if the rules of superselection are taken into account.

**Acknowledgement.** The author appreciates greatly A.V. Grigoriev and S.I. Svertilov for the fruitful discussion of the article.

# References

[1] Pontecorvo B. 1957 Zh. Eksp.Teor.Fiz.**34** 247

   [1958 Sov.Phys.JETP **7** 172]

[2] Maki Z., Nakagawa M., Sakata S. 1962 Prog. Theor. Phys. **28** 870

[3] Gribov V.,Pontecorvo B. 1969 Phys. Lett. B **28** 493

[4] Wolfenstein L. 1978 Phys. Rev. D **17** 2369

[5] Mikheev S., Smirnov A. 1985 Yad. Fiz. **42** 1441

   [Sov. J. Nucl. Phys. **42** 913]

[6] Fock V., Krylov S. 1947 Zh. Eksp. Teor. Fiz. **17** 93

[7] Kayser B. 1981 Phys. Rev. D **24** 110

[8] Blatt J., Weisskopf V. Theoretical nuclear physics,

   New-York-London (1952)

[9] Davydov A. Quantum Mechanics, Pergamon Press (1965)

[10] Baz A., Zel`dovich Ya., Perelomov A. Scattering, reactions and decay

   in nonrelativistic quantum mechanics, Jerusalem (1969)

[11] Wick G., Wigner E., Wightman A. 1952 Phys. Rew. **88** 101;1970 D **1** 3267

# Appendix

Usually within 3$f$-variant mixing matrix $U$ with following designations is used:

$$u_{a1} = c_{13}c_{12} \quad u_{a2} = c_{13}s_{12} \quad u_{a3} = s_{13} \tag{A.1.1}$$

$$u_{b1} = -c_{23}s_{12} - s_{23}s_{13}c_{12} \quad u_{b2} = c_{23}c_{12} - s_{23}s_{13}s_{12} \quad u_{b3} = s_{23}c_{13} \tag{A.1.2}$$

$$u_{c1} = s_{23}s_{12} - c_{23}s_{13}c_{12} \quad u_{c2} = -s_{23}c_{12} - c_{23}s_{13}s_{12} \quad u_{c3} = c_{23}c_{13} \tag{A.1.3}$$

Here symbols $s$ and $c$ mean sine and cosine functions, respectively.

Matrix (A.1) corresponds to the case of *CP*-invariance conservation.

In this article more convenient and compact expressions for mixing matrix elements are used:

$$u_{a1} = c_2c_3 \quad u_{a2} = c_2s_3 \quad u_{a3} = s_2 \tag{A.2.1}$$

$$u_{b1} = -c_1s_3 - s_1s_2c_3 \quad u_{b2} = c_1c_3 - s_1s_2s_3 \quad u_{b3} = s_1c_2 \tag{A.2.2}$$

$$u_{c1} = s_1s_3 - c_1s_2c_3 \quad u_{c2} = -s_1c_3 - c_1s_2s_3 \quad u_{c3} = c_1c_2 \tag{A.2.3}$$

Matrix (A.2) can be reduced to the mixing matrix of 2$f$-variant, if we set $\theta_2 = 0$ (i.e. $s_2 = 0$ and $c_2 = 1$) and $\theta_1 = 0$ ($s_1 = 0$ and $c_1 = 1$). In this case only one mixing angle $\theta_3$ remains (see formulas 2a and 2b).